\documentclass [aps,amsmath,amssymb,superscriptaddress,preprint]{revtex4}
\usepackage{graphicx}

\newcommand{\nc}{\newcommand}
\nc{\beq}{\begin{equation}}
\nc{\eeq}{\end{equation}}
\nc{\beqa}{\begin{eqnarray}}
\nc{\eeqa}{\end{eqnarray}}

\def\gsim{\mathrel{\rlap{\lower4pt\hbox{\hskip1pt$\sim$}}
    \raise1pt\hbox{$>$}}}       

\begin{document}

\title{The measure problem in no-collapse (many worlds) quantum mechanics}  

\author{Stephen~D.~H.~Hsu} \email{hsu@msu.edu}
\affiliation{Department of Physics and Astronomy\\ Michigan State University}

\begin{abstract}
We explain the measure problem (cf. origin of the Born probability rule) in no-collapse quantum mechanics. Everett defined {\it maverick} branches of the state vector as those on which the usual Born probability rule fails to hold -- these branches exhibit highly improbable behaviors, including possibly the breakdown of decoherence or even the absence of an emergent semi-classical reality. Derivations of the Born rule which originate in decision theory or subjective probability (i.e., the reasoning of individual observers) do not resolve this problem, because they are circular: they assume, a priori, that the observer occupies a non-maverick branch. An ab initio probability measure is sometimes assumed to explain why we do not occupy a maverick branch. This measure is constrained by, e.g., Gleason's Theorem or envariance to be the usual Hilbert measure. However, this ab initio measure ultimately governs the allocation of a self or a consciousness to a particular branch of the wavefunction, and hence invokes primitives which lie beyond the Everett wavefunction and beyond what we usually think of as physics. The significance of this leap has been largely overlooked, but requires serious scrutiny.
\end{abstract}


\maketitle

\date{today}


\section{Do wave functions collapse?}

We begin with a brief introduction to no-collapse or many worlds quantum mechanics. A longer version of this material can be found in an earlier paper by the author \cite{Hsu-prob}. Readers already familiar with this subject can skip to section II.

Quantum mechanics, as conventionally formulated, has two types of time evolution. An isolated system $S$ evolves according to the Schrodinger equation
\begin{equation}
\label{unitary}
\vert \Psi (t) \rangle = \exp ( -i H t )~ \vert \Psi (0) \rangle ~,
\end{equation}
where the time evolution operator $U(t) = \exp ( -i H t )$ is unitary: $U^\dagger U = 1$. However, when a measurement is made the state undergoes non-unitary von Neumann projection (i.e., collapses) to an eigenstate corresponding to the eigenvalue observed by the measuring device. The two types of time evolution are so radically different that a rigorous definition of exactly when each of them apply is essential. Unfortunately, as is widely acknowledged, the conventional interpretation does not supply a satisfactory definition -- see, e.g., {\it Against Measurement} by J.S. Bell \cite{Bell}.

Can the measurement process be described within a larger isolated system $S'$, containing both the original system $S$ and the measuring apparatus $M$, whose dynamics is still unitary? Crucially, measurement collapse and unitary evolution of $S'$ are incompatible, because the projection operation is not invertible. Collapse is fundamentally irreversible, whereas unitary evolution of the system and measuring device together is reversible. Nevertheless, as was first noted by Everett \cite{Everett, Everett1}, unitary evolution of the larger system $S'$ is compatible with the {\it appearance} of collapse to observers within the system.

To make this discussion more explicit, let $S$ be a single qubit and $M$ a device (e.g., Stern-Gerlach device) which measures the spin of the qubit along a particular axis. The eigenstates of spin along this axis are denoted $\vert \pm \rangle$. We define the measuring process as follows (the arrow denotes time evolution):
\begin{eqnarray}
\vert + \rangle \otimes \vert M \rangle ~\longrightarrow ~  \vert + \, , \, M_+ \rangle  \nonumber \\
\vert - \rangle \otimes \vert M \rangle ~\longrightarrow ~  \vert - \,,\, M_- \rangle 
\end{eqnarray}
where $M_+$ denotes a state of the apparatus which has recorded a $+$ outcome, and similarly with $M_-$. We can then ask what happens to a superposition state $\vert \Psi_S \rangle =  c_+ \vert + \rangle ~+~ c_- \vert - \rangle$ which enters the device $M$. In the conventional formulation, with measurement collapse, {\it one} of the two final states $\vert + \, , \, M_+ \rangle~$ {\it or} $~\vert - \, , \, M_- \rangle$ results, with probabilities $\vert c_+ \vert^2$ and $\vert c_- \vert^2$ respectively. This probability rule is called the Born rule. In the conventional formulation the notion of probability enters precisely when one introduces wave function collapse. 

However, if the combined system $S' = S + M$ evolves unitarily (in particular, {\it linearly}) as in (\ref{unitary}), we obtain a superposition of measurement device states (see figure 1):
\begin{equation}
\big(  c_+ \vert + \rangle ~+~ c_- \vert - \rangle  \big ) \otimes \vert M \rangle ~~ \longrightarrow ~~
c_+ \, \vert + \, , \, M_+ \rangle ~+~ c_- \, \vert - \, , \, M_- \rangle ~~.
\end{equation}
This seems counter to actual experience: properly executed measurements produce a single outcome, not a superposition state. But how do we know for sure? In fact, an observer described by the state $M_+$ might be unaware of the second branch of the wave function in state $M_-$ for dynamical reasons related to a phenomenon called {\it decoherence} \cite{decoherence,decoherence1,decoherence2,decoherence3}. First, any object sufficiently complex to be considered either a measuring device or observer (i.e., which, in the conventional formulation can be regarded as semi-classical) will have many degrees of freedom. Second, a measurement can only be said to have occurred if the states $M_+$ and $M_-$ differ radically: the outcome of the measurement must be stored in a redundant and macroscopically accessible way in the device (or, equivalently, in the local environment). Therefore, the ability of the second branch to affect the first branch is exponentially small: the overlap of $M_+$ with $M_-$ is typically of order $\exp( - N )$, where $N$ is a macroscopic number of degrees of freedom, and the future dynamical evolution of each branch is unlikely to alter this situation. For All Practical Purposes (FAPP, as J.S. Bell put it), an observer on one branch can ignore the existence of the other: they are said to have decohered. Each of the two observers will perceive a collapse to have occurred, although the evolution of the overall system $S'$ has continued to obey the Schrodinger equation. 

It is sometimes said that decoherence solves the measurement problem in quantum mechanics. More accurately, decoherence illuminates the process of measurement. But it does not answer the question of whether or not wave functions actually collapse -- {\it it merely makes it clear that they need not!} Measurement can proceed in a perfectly continuous manner, with the different branches rapidly losing contact with each other. The result is the {\it appearance} of a single outcome (to observers $M_{\pm}$ within the system $S'$) without invoking collapse. A no-collapse universe is an intrinsically quantum object, containing all possible amplitudes, but in which (semi-classical) observers with discordant memory records are unaware of each others' existence.

Without collapse the status of probability is much more subtle. We explore this issue in detail in the next section.

\bigskip

\section{Probability and deterministic evolution}

The Schrodinger dynamics governing unitary evolution of the wave function is entirely deterministic. In the absence of collapse, this is the only kind of time evolution in quantum mechanics. If the initial state of $S'$ is known at some initial time $t_0$, it can be predicted with perfect accuracy at any subsequent time $t$. How, in such a deterministic system, can the notion of probability arise?

In the conventional interpretation, with only a single realized outcome of an experiment, one simply {\it imposes} the Born probability rule together with measurement collapse: the probability of the plus outcome is $|c_+|^2$, or more generally the likelihood associated with a particular component of the wave function is given by its magnitude squared under the Hilbert measure. Philosophically, this imposition of objective randomness is a violent departure from the notion that all things should have causes: in the conventional (Copenhagen) interpretation of quantum mechanics there is {\it no cause} for the specific outcome that is observed. This is what is meant by real or objective randomness.

In the absence of collapse there is no logical point at which we can introduce a probability rule -- it must emerge by itself, and it must explain why an experimenter, who decoheres into different versions of himself on different branches, does so with probabilities determined by $\vert c_\pm \vert^2$. When all outcomes are realized, what is meant by the {\it probability} of a particular one? 

The problem is actually worse than this. For decoherence to work in the first place, the overlap or Hilbert space inner product between two vectors must have a probabilistic interpretation: if the inner product is small (decoherence has occurred) the two branches are assumed not to influence each other; an observer on one branch may be unaware of the other. Let us leave this further, more subtle, complication for the next section and here simply assume that decoherence and the measuring apparatus work as desired. Even in this reduced context, there is still a problem, as we now discuss.

The many worlds interpretation must provide a probability measure over all decoherent outcomes, each of which is realized, i.e., from the perspective of at least one observer or recording device. The difficulty becomes clear if we consider $N$ spins: $\Psi = \otimes_{i=1}^N ~ \psi_i$, with each spin prepared in the identical state $\psi_i = c_+ \vert + \rangle_i ~+~ c_- \vert - \rangle_i$. Again, all possibilities are realized, including the outcome where, e.g., all spins are measured in the $+$ state: $\Psi \sim \vert + + + \cdots + \rangle$. If $\vert c_+ \vert$ is small, then this outcome is very unlikely according to the usual probability rules of quantum mechanics. However, independent of the value of $c_+$, it comprises one of the $2^N$ distinct possible outcomes. Each of these outcomes implies the existence of an observer with distinct memory records. 

For $N$ sufficiently large and $|c_+| \neq |c_-|$, it can be shown \cite{BHZ} that the vast majority of the $2^N$ realized observers (i.e., counting each distinct observer equally) see an outcome which is highly unlikely according to the usual probability rules. Note that counting of possible outcomes depends only on combinatorics and is independent of $c_\pm$. As $N \rightarrow \infty$, for all values of $c_\pm$ (excluding exactly zero and one), almost all of the realized observers find nearly equal number of $+$ and $-$ spins: there are many more outcomes of the form, e.g., $(+ + - + - \, \cdots \, + - - +)$ with roughly equal number of $+$'s and $-$'s than with many more of one than the other. This had to be the case, because counting of outcomes is independent of the values of $c_\pm$, leading to a symmetry between $+$ and $-$ outcomes in the combinatorics. In contrast, the Born rule predicts that the relative number of $+$ and $-$ outcomes depends on $|c_\pm|^2$. In the large $N$ limit almost all (distinct) observers \cite{FN1} experience outcomes that strongly disfavor the Born probability rule: {\it almost all of the physicists in the multiverse see experimental violation of the Born rule}. Or: {\it a vanishing fraction of the physicists in the multiverse see outcomes consistent with the Born rule.} In using terms like {\it almost all}, we have momentarily adopted a specific measure: the simple {\it counting measure} in which each of the $2^N$ observers is weighted equally. While we do not mean to advocate for this measure over others, it provides a simple illustration of how essential the choice of measure is to the interpretation of the theory.

Everett referred to the branches on which results deviate strongly from Born rule predictions (i.e., exhibit highly improbable results according to the usual probability rule) as {\it maverick} branches. By definition, the magnitude of these components under the Hilbert measure vanishes as $N$ becomes large. But there is no sense in which the Hilbert measure is privileged in many worlds. Nor is there even a logical place to introduce it -- it must emerge in some way by itself. Everett claimed to derive quantum mechanical probability by taking $N$ to infinity and discarding all zero norm states in this limit, thereby eliminating all maverick outcomes. Most advocates of many worlds regard this reasoning as circular and look elsewhere for a justification of the Born rule.

Instead, most attempts to justify the Born rule have relied on {\it subjective} probability arguments. (Dynamical mechanisms for removing maverick branches have also been considered \cite{BHZ}.) While objective probabilities can be defined through frequencies of outcomes of a {\it truly random} process, subjective probabilities deal with degrees of belief. The conventional quantum interpretation, with von Neumann projection, assumes true randomness and objective probabilities: repeated measurements produce a truly random sequence of outcomes, with frequencies given by the Born rule. Outcomes are unknowable, {\it even in principle}, even with perfect knowledge of the state being measured. (Einstein objected to the introduction of true randomness, because it implies outcomes without causes.) In the absence of true randomness, such as in classical (deterministic) physics, individuals with limited knowledge reason using subjective probabilities, which represent degrees of belief \cite{FN2}.

Let us elaborate on the sense in which no collapse quantum mechanics can be considered deterministic. Just before the measurement, all of the observers are in identical states. Using the basis $\vert m \rangle \equiv \vert s_1, s_2, \cdots s_N \rangle$, where $s_i$ are individual spin eigenstates and $m$ runs from $1$ to $2^N$, we can write
\begin{eqnarray}
\label{copies}
\sum_{m}  c_m \, \vert m \rangle \otimes \vert {\cal O} \rangle ~ &&= ~ c_1 \, \vert 1 \rangle \otimes \vert {\cal O} \rangle
~+~ c_2 \, \vert 2 \rangle \otimes \vert {\cal O} \rangle ~+~ c_3 \, \vert 3 \rangle \otimes \vert {\cal O} \rangle  ~+~ \cdots    \\
 && \longrightarrow ~~   c_1 \, \vert 1 \, , \, {\cal O}_1 \rangle
~+~ c_2 \, \vert 2 \, , \, {\cal O}_2 \rangle ~+~ c_3 \, \vert 3 \, , \, {\cal O}_3 \rangle  ~+~ \cdots ~=~ \sum_{m}  c_m \, \vert m \, , \, {\cal O}_m \rangle~~ \nonumber
\end{eqnarray}
where ${\cal O}_m$ denotes the state describing the observer who recorded outcome $m$. The first line has been written to emphasize that in the $m$ basis it appears as if there are $2^N$ identical observers, each of whom is {\it destined} to evolve into a particular one of the ${\cal O}_m$. Of course, the observer does not know which of the ${\cal O}_m$ they will evolve into, because they do not (yet) know the spin state $\vert m \rangle$ on their branch. But the outcome can be considered pre-determined: it was caused by the value of $m$ on the branch of that observer \cite{FN3}. This perspective may appear more natural if one considers the time reversal of the final state in (\ref{copies}). Each observer ${\cal O}_m$ evolves backward in time to one of the identical $\cal O$'s. 

Of course, this still neglects the question of why {\it my consciousness} in particular has been assigned to a specific decoherent branch of the universal wave function. A measure which is imposed {\it ab initio} -- which allocates a self or consciousness to a specific branch $m$ with a certain probability -- does provide a means to introduce objective probability and true randomness into this discussion, but at the cost of also having to introduce something far beyond the universal wave function $\Psi$. The notion of a self or a consciousness that exists outside $\Psi$ but is then associated with a particular component of $\Psi$ with some probability is, to this author, a last resort that should be avoided if possible. See Figure 1.

\begin{figure}[t!]
\label{qm}
\includegraphics[width=12cm,height=6cm]{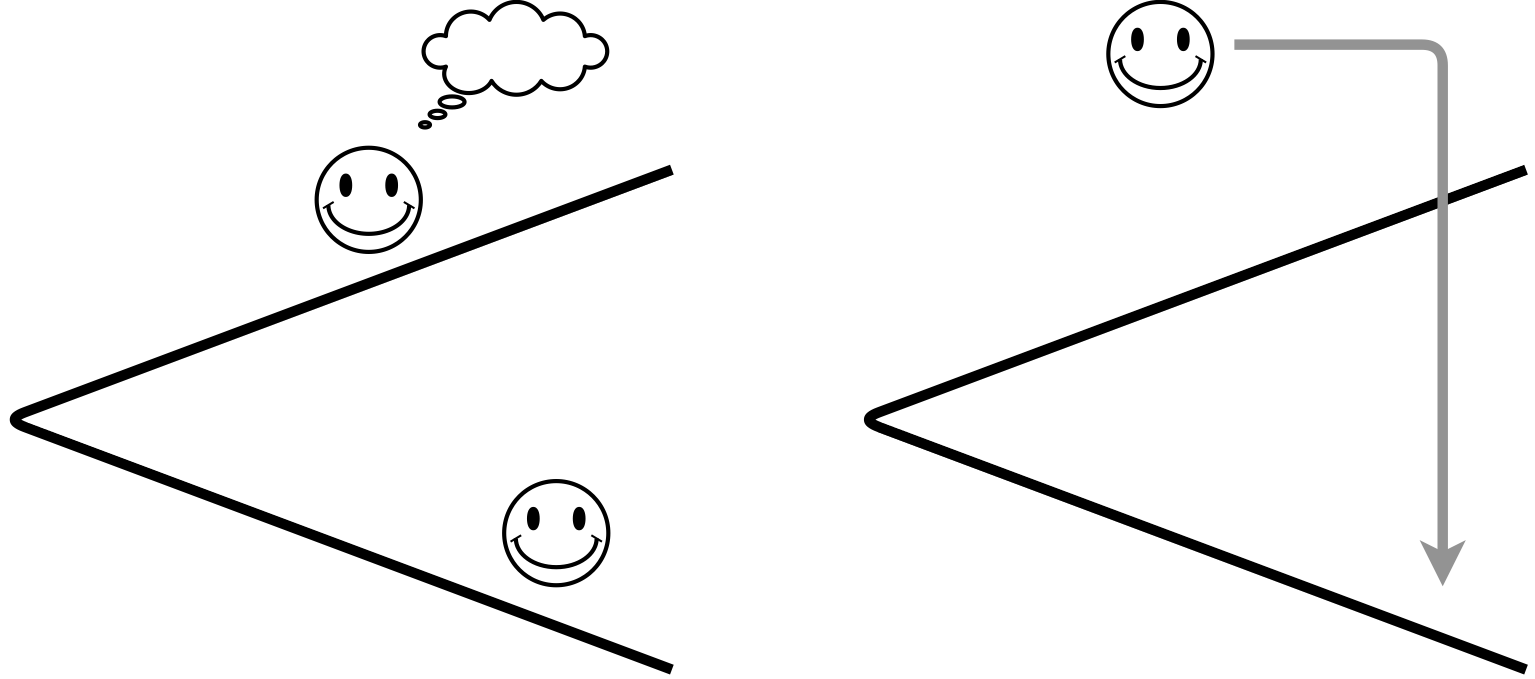}
\caption{Two types of probability. Left: Subjective probability, or degrees of belief (reasoning) of an observer. Right: Objective probability, governing the assignment of a self to a particular branch.}
\end{figure}

Subjective probability arguments originate in certain postulates governing the way in which we {\it reason} about probabilities \cite{Deutsch,Wallace,Sean}. For example, following Laplace, we might require that two components of the wave function related by symmetry (i.e., with equal coefficients $c_i$ and $c_j$) must have equal probabilities as outcomes. By further analysis, we might conclude that probability must be proportional to Hilbert space norm squared. The reader can judge for himself whether these arguments are convincing \cite{FN4,Dawid}. It is important to emphasize that, even if successful, subjective probability arguments do not alter the fact that the many worlds wave function realizes vast numbers of observers (all but a vanishing fraction, under the counting measure discussed earlier) who see gigantic violations of the Born rule (even, gross departures from decoherence). This vast number of observers seem unlikely to believe in conventional quantum mechanics as the correct theory of Nature, let alone to conform to the reasoning described above. We explore this issue in the following section.

\bigskip

\section{The measure problem unresolved}

In the previous section we illustrated the drastic differences resulting from two choices of measure: the counting distinct histories measure vs the Hilbert (norm-squared) measure. The former leads to predominance of maverick branches, the latter to the usual quantum mechanics and the Born rule. Obviously, there is an infinite set of possible measures over the space of state vectors. Yet, there is not even a natural place in the theory to impose a measure. Everett claimed (erroneously) that the Hilbert measure emerges naturally, but in fact the measure problem remains unsolved. 

Let us generalize from the $N$ spin measurement discussed in the previous section to a no-collapse wavefunction describing an entire universe. For simplicity, consider a closed (finite) universe with a finite number of degrees of freedom, so that the dimensionality of the Hilbert space, while large, is not infinite. Consider all the possible histories produced by the evolution of the wavefunction. These must include those in which {\bf (A)} the Born rule is strongly violated, and {\bf (B)} decoherence is violated. We discussed {\bf (A)} in detail already, but clearly decoherence itself is a consequence of apparent (statistical) irreversibility in the function of measurement devices, and need not hold on all histories. (That decoherent branches are unlikely to re-cohere or influence each other is itself a probabilistic statement based on the Born rule.) Furthermore, once we allow deviation from the usual quantum probabilities, histories with {\bf (C)} no emergent semi-classical reality are possible.

Because this paper is about the Everett interpretation, we simply assume that ordinary quantum dynamics together with the Born rule lead to decoherent outcomes and an emergent semi-classical reality consistent with our experience. Critics of many worlds are right to emphasize that this is an unproven conjecture.  On maverick branches, which deviate strongly from Born probabilities, there is no reason to expect that these properties hold, and not even Everett proponents suggest that they do.

Histories with properties {\bf (A)}, {\bf (B)}, or {\bf (C)} are nearly zero measure if the usual Hilbert measure is assumed {\it ab initio} -- i.e., imposed by hand. However, under other measures, like the distinct history counting measure \cite{FN1} discussed previously, maverick histories of the type {\bf (A-C)} dominate. Subjective probability discussions such as \cite{Deutsch, Sean} all {\it assume} a non-maverick branch, and {\it then} discuss the reasoning or decision theoretic considerations of observers on that branch. However, the initial assumption cannot be justified without use of the Hilbert measure, or some other reason for excluding the maverick branches. (See \cite{BHZ} for a specific proposal.) The question which is not addressed by subjective probability discussions is {\it why} we (you and I) ended up on a non-maverick branch of the universal wavefunction in the first place. It is logically independent of arguments about how we (you and I) should reason, given that our memory records are consistent with the Born rule, decoherence, and a semi-classical reality. In fact, the question to be resolved is similar to the type that arises in any theory of a multiverse (e.g., the string theory landscape): What explains the atypical (relative to other universes) features of our world? 

Is there, perhaps, an anthropic justification for excluding maverick histories? For example, is it possible that information-processing observers are highly unlikely to arise when any of {\bf (A-C)} apply? This seems implausible, because significant deviations from the usual quantum probabilities do not seem to preclude complex life. For example, suppose that decoherence were to operate orders of magnitude more slowly than in usual quantum dynamics, because of large (improbable) fluctuations in measuring devices or the environment. But decoherence timescales are many, many orders of magnitude smaller than the relevant timescales for biological processes. So, slower decoherence would not hinder life, despite making this type of branch highly unusual under the Hilbert measure. Also, brain function appears to be essentially classical. Larger deviations from semi-classicality (i.e., additional randomness beyond the usual biophysics) might be a problem that requires additional error-correction, but does not seem to be catastrophic to intelligence. In contrast, if there were an anthropic principle excluding maverick branches, we would expect complex life to be very sensitive to deviations from usual quantum dynamics.

Without a mechanism for excluding maverick branches, we cannot explain why observers living in a no-collapse quantum universe should find that the Born rule for probabilities holds, decoherence operates under the usual quantum dynamics, and a semi-classical reality emerges from quantum mechanics. This is the (unresolved) measure problem in many worlds quantum mechanics.

One way to summarize this discussion is to think about a no-collapse universe as a collection of many minds: each distinct observer, with his distinct memory records, is one of the minds. Suppose that I, writing this paper, and you, reading it, share the same branch of the wavefunction. Why should it not be a maverick branch? It seems that this outcome requires either the {\it ab initio} imposition of a measure (i.e., the Hilbert measure), or some other means (e.g., dynamical \cite{BHZ}) of eliminating the most egregious mavericks. It is not surprising that a subjective probability analysis which takes place on a non-maverick branch reproduces the Born rule, because the history of that branch, the functioning of measurement devices, etc., are assumed {\it a priori} to obey the Born rule. Rational observers who extrapolate the past into the future will make wagers or decisions according to this rule.

\bigskip

\section{Envariance and Gleason's Theorem}

Finally, let us discuss Gleason's Theorem \cite{gleason} and Zurek's envariance arguments \cite{zurek}.
Zurek has proposed an ingenious derivation of the Born rule based on envariance, defined as environmentally assisted invariance or invariance of entangled states under simultaneous unitary transformation of system and environment. Envariance arguments are similar to Gleason's Theorem, in that minimal assumptions are shown to lead to the Born rule. Like Gleason's Theorem, envariance arguments can be applied either to subjective probabilities (but, as discussed in the previous section, only to the reasoning of individuals on non-maverick branches, due to {\bf (A-C)}) or to objective, {\it ab initio} probability measures that are imposed by hand. 

In the case of an {\it ab initio} measure (see Figure 1, right side), one still has to explain the meaning of the probabilities. The probability has to do with the uncertainty that an observer has, pre-measurement, concerning which branch they will occupy post-measurement. In equation (\ref{copies}), this uncertainty corresponds to which $m$ branch the observer is on before the measurement. The measure determines the likelihood that a specific consciousness or mind or self is located on or assigned to a particular $m$ branch. This likelihood is not intrinsic to the universal wavefunction or its Schrodinger evolution and has to be imposed from outside. This is in contrast to Everett's original hope that the ``theory would provide its own interpretation'' \cite{Everett}.

\bigskip

\section{Conclusion}

In a 2005 Physics Today article entitled {\it Einstein's Mistakes} \cite{Weinberg}, Steven Weinberg wrote  

\begin{quote}
Bohr's version of quantum mechanics was deeply flawed, but not for the reason Einstein thought. The Copenhagen interpretation describes what happens when an observer makes a measurement, but the observer and the act of measurement are themselves treated classically. This is surely wrong: Physicists and their apparatus must be governed by the same quantum mechanical rules that govern everything else in the universe. But these rules are expressed in terms of a wavefunction (or, more precisely, a state vector) that evolves in a perfectly deterministic way. So where do the probabilistic rules of the Copenhagen interpretation come from? 

The Copenhagen rules clearly work, so they have to be accepted. But this leaves the task of explaining them by applying the deterministic equation for the evolution of the wavefunction, the Schrodinger equation, to observers and their apparatus. The difficulty is not that quantum mechanics is probabilistic ... The real difficulty is that it is also deterministic, or more precisely, that it combines a probabilistic interpretation with deterministic dynamics. ...
\end{quote}

Proposed solutions to Weinberg's difficulty which are based in decision theory or subjective probability (i.e., the reasoning of individuals on different branches) do not fully harmonize the deterministic evolution of the wavefunction with our everyday experience.  The universal wavefunction contains vast numbers of maverick branches -- on which the Born rule, decoherence, or emergence of classical reality are violated \cite{FN1}. What is missing is an explanation for {\it why} we (you and I) do not inhabit a maverick branch. Arguments based on rationality, but which describe the reasoning of an observer in a non-maverick world, cannot resolve this question.

It is sometimes assumed that one can simply impose an {\it ab initio} probability measure by hand, which is then constrained by Gleason's theorem or envariance to be the Born measure. However, as has been emphasized several times in this paper, the nature of this measure needs to be considered more carefully. This measure governs something that we might call ``self-locating uncertainty'' -- i.e., the assignment of a mind or consciousness to specific branches of the wavefunction.  But what exactly is this ``self''? It seems to be something that exists outside of quantum mechanics and the usual physical degrees of freedom. (See Figure 1, right side)

As a concrete example, imagine an alien being playing a computer game that takes place in his desktop quantum computer. The alien's consciousness is projected into the game, so that he experiences exactly one branch among the many that are realized in the quantum superposition state $\Psi$. The objective probability measure that is often casually invoked in many worlds quantum mechanics governs which branch the alien mind is allocated to. The alien must exist outside of $\Psi$, and hence {\it outside the usual realm of physics}. Surely we can hope for an explanation for the Born rule in quantum mechanics without this taking this kind of leap.

Return again to Weinberg, this time his lectures on quantum mechanics \cite{Wtext}:
\begin{quote}
... it would be disappointing if we had to give up the ``realist'' goal of finding complete descriptions of physical systems, and of using this description to derive the Born rule, rather than just assuming it. We can live with the idea that the state of a physical system is described by a vector in Hilbert space rather than by numerical values of the positions and momenta of all the particles in the system, but it is hard to live with no description of physical states at all, only an algorithm for calculating probabilities. My own conclusion (not universally shared) is that today {\it there is no interpretation of quantum mechanics that does not have serious flaws} [italics added] ...
\end{quote}

\bigskip

\emph{Acknowledgements---} The author thanks Jess Riedel and Steve Avery for useful comments. This work was supported in part by the Office of the Vice-President for Research and Graduate Studies at MSU. 


\bigskip



\begin{thebibliography}{99}
\bibitem{FN1} If one adopts the simple counting measure over discrete histories, then almost all observers exist on maverick branches.
\bibitem{FN3} The {\it reason} that observer ${\cal O}_m$ observed outcome $m$ is that, before the measurement, he was paired with the spin state $\vert m \rangle$: $~\, \vert m \rangle \otimes \vert {\cal O} \rangle ~\rightarrow~ \vert m \, , \, {\cal O}_m \rangle$. This initial condition {\it caused} the measurement outcome. Compare to the conventional formulation, in which, e.g., the observer measuring spin state $( \vert + \rangle \, + \, \vert - \rangle )$ finds the outcome $+$ and the wave function collapses to $\vert + \rangle$ only, with no $\vert - \rangle$ branch. Clearly there is no {\it cause} for the $+$ outcome (as opposed to $-$): the result is truly, objectively random, something that Einstein objected violently to.
\end{thebibliography}
\end{document}